\RequirePackage{fix-cm}
\documentclass[twocolumn]{svjour3}          % twocolumn
\smartqed  % flush right qed marks, e.g. at end of proof
\usepackage{mathptmx}      % use Times fonts if available on your TeX system
\usepackage{graphicx,color} %png
\usepackage{amsmath}
\usepackage{epstopdf}
\usepackage{textgreek}
\usepackage{amssymb}
\usepackage{caption}
\usepackage{ulem}
\usepackage[english]{babel}
\newcommand{\um}{\textmu m\,}
\newcommand{\degree}{\ensuremath{^\circ}\,}
\newcommand{\minute}{\ensuremath{^\prime}\,}
\newcommand{\second}{\ensuremath{^{\prime\prime}}\,}

\newcommand{\Planck}{{\it Planck}}  % 
\newcommand{\Herschel}{{\it Herschel}}  % 
\newcommand{\Pacs}{{\it PACS}}  %
\newcommand{\Estadius}{{\it Estadius}}  %
\newcommand{\ACTPol}{{\it ACTPol}}  % 
\newcommand{\Pilot}{{\it PILOT}}  % 
\newcommand{\PILOT}{{\it PILOT}}  % 
\newcommand{\Blastpol}{{\it BLASTPol}}  % 
\newcommand{\HAWCplus}{{\it HAWC+}}  % 
\newcommand{\SCUBApol}{{\it SCUBA-pol}}  % 

\newcommand{\SOFIA}{{\it SOFIA}}  % 
\newcommand{\JCMT}{{\it JCMT}}  % 
\newcommand{\IRAM}{{\it IRAM}}  % 

\newcommand{\ICS}{ICS}  % 
\newcommand{\HWP}{HWP}  % 

\newcommand{\leakage}{leakage}  % 
\newcommand{\Leakage}{Leakage}  % 
\newcommand{\crosstalk}{cross-talk}  % 
\newcommand{\Crosstalk}{Cross-talk}  % 
\newcommand{\ghost}{read-out~latency}  % 
\newcommand{\Ghost}{Read-out~latency}  % 

\newcommand{\fct}{f_{CT}} % crosstalk coeficient
\newcommand{\frl}{f_{RL}} % Readout Latency coeficient

\newcommand{\scanamorphos}{Scanamorphos} %GdG
\newcommand{\NIKAtwo}{\it NIKA2} %GdG

\newcommand{\hwppos}{HWP_{\rm pos}}  %HWP position

\newcommand{\mic}{\,{\rm \mu m} }
\newcommand{\degr}{{\rm ^o} }

\newcommand{\TRANS}{TRANS}  %Transmission array
\newcommand{\REFLEX}{REFLEX}  %Reflexion array
\newcommand{\flightthree}{flight\#3}

\newcommand{\iprime}{k}

\newcommand{\StokesI}{I}                    %Intensity
\newcommand{\StokesQ}{Q}                    %Q
\newcommand{\StokesU}{U}                    %U
\newcommand{\polfrac}{p}                    %polarization fraction, not degree. 
\newcommand{\polang}{\psi}                  %polarization angle (CMB convention)

\begin{document}

\title{Performance of the polarization leakage correction in the PILOT data
\thanks{The authors acknowledge support from the ballooning division at CNES.
%Grants or other notes about the article that should go on the front page should be placed here. General acknowledgments should be placed at the end of the article.
}
}

\titlerunning{The PILOT polarization leakage correction}

\author{
J-Ph. Bernard,
A. Bernard,
H. Roussel,
I. Choubani,
D. Alina,
J. Aumont,
A. Hughes,
I. Ristorcelli,
S. Stever,
T. Matsumura
S. Sugiyama,
K. Komatsu,
G. de~Gasperis,
K. Ferrière,
V. Guillet,
N. Ysard,
%%%%%%%%%%%%%%% start of alphabetical
P. Ade,
P. de~Bernardis,
N. Bray,
B. Crane,
J.P. Dubois,
M. Griffin,
P. Hargrave,
Y. Longval,
S. Louvel,
B. Maffei,
S. Masi,
B. Mot,
J. Montel,
F. Pajot,
E. P\'erot,
N. Ponthieu,
L. Rodriguez,
V. Sauvage,
G. Savini,
C. Tucker,
F. Vacher
}

\authorrunning{J.-Ph. Bernard et al.} % if too long for running head

\institute{
J-Ph. Bernard,
A. Bernard,
I. Chubani,
J. Aumont,
A. Hughes,
I. Ristorcelli,
K. Ferrière,
B. Mot,
F. Pajot 
 \at
Institut de Recherche en Astrophysique et Planetologie
(IRAP), Universit\'e Paul Sabatier,
9 Av du Colonel Roche, BP 4346, 31028 Toulouse cedex 4; \\
\email{Jean-Philippe.Bernard@irap.omp.eu}           %  \\
%%%%%%%%%%%%%
\and
D. Alina
\at
Nazarbayev University, Physics Department
53 Kabanbay Batyr Avenue, Nur-Sultan city, Kazakhstan
 \\
%%%%%%%%%%%%%
\and
S. Sugiyama,
K. Komatsu,
T. Matsumura
\at
Kavli Institute for the Physics and Mathematics of the Universe (Kavli IPMU)(WPI)
The University of Tokyo Institutes for Advanced Study (UTIAS)
The University of Tokyo
5-1-5 Kashiwa-no-Ha, Kashiwa City, Chiba 277-8583, Japan
 \\
\and
S. Stever,
\at
Kavli Institute for the Physics and Mathematics of the Universe (Kavli IPMU)(WPI)
The University of Tokyo Institutes for Advanced Study (UTIAS)
The University of Tokyo
5-1-5 Kashiwa-no-Ha, Kashiwa City, Chiba 277-8583, Japan\\
Okayama University,  3-1-1, Tsushimanaka, Kita-ku, Okayama City, Okayama, 700-8530, Japan \\
%%%%%%%%%%%%%
\and
B. Maffei,
V. Guillet,
N. Ysard,
B. Crane,
J-P. Dubois,
Y. Longval,
V. Sauvage
\at
Institut d’Astrophysique Spatiale, CNRS, Université Paris-Saclay, Bât.121, 91405 Orsay cedex, France
 \\
%%%%%%%%%%%%%
\and
P. de~Bernardis,
S. Masi
\at
Universit\`a degli studi di Roma ``La Sapienza'', Dipartimento di Fisica, 
P.le A. Moro, 2, 00185, Roma, Italia; \\
%%%%%%%%%%%%%
\and
P. Ade,
M. Griffin,
P. Hargrave,
G. Savini,
C. Tucker
\at
Department of Physics and Astrophysics, PO BOX 913, Cardiff University, 
5 the Parade, Cardiff, UK; \\
%%%%%%%%%%%%%
\and
G. de~Gasperis
\at
Universit\`a degli studi di Roma ``Tor Vergata'', Dipartimento di Fisica, V.le della Ricerca Scientifica, 1, 00133 Roma, Italia;\\
%%%%%%%%%%%%%
\and
L. Rodriguez
\at
CEA/Saclay, 91191 Gif-sur-Yvette Cedex, France; \\
%%%%%%%%%%%%%
\and
N. Ponthieu
\at
Univ. Grenoble Alpes, CNRS, IPAG, 38000 Grenoble, France; \\
%%%%%%%%%%%%%
\and
H. Roussel
\at 
Institut d'Astrophysique de Paris, Sorbonne Université, CNRS (UMR 7095),
98 bis boulevard Arago, 75014 Paris, France;\\
%%%%%%%%%%%%%
\and
N. Bray,
S. Louvel,
J. Montel,
E. P\'erot,
F. Vacher
\at
Centre National des Etudes Spatiales, DCT/BL/NB, 18 Av. E.
Belin, 31401  Toulouse, France; \\
}

\date{Received: date / Accepted: date}

\maketitle

\begin{abstract}

The Polarized Instrument for Long-wavelength\sloppy\ Observation of the Tenuous interstellar medium ({\Pilot}) is a balloon-borne experiment that aims to measure the polarized emission of thermal dust at a wavelength of 240\,\um (1.2\,THz). The {\Pilot} experiment flew from Timmins, Ontario, Canada in 2015 and 2019 and from Alice Springs, Australia in April 2017. The in-flight performance of the instrument during the second flight was described in \cite{Mangilli2019a}.
In this paper, we present data processing steps that were not presented in \cite{Mangilli2019a} and that we have recently implemented to correct for several remaining instrumental effects. The additional data processing concerns corrections related to detector cross-talk and readout circuit memory effects, and leakage from total intensity to polarization.
We illustrate the above effects and the performance of our corrections using data obtained during the third flight of {\PILOT}, but the methods used to assess the impact of these effects on the final science-ready data, and our strategies for correcting them will be applied to all {\PILOT} data.
We show that the above corrections, and in particular that for the intensity to polarization leakage, which is most critical for accurate polarization measurements with {\PILOT}, are accurate to better than 0.4\% as measured on Jupiter during {\flightthree}.

\end{abstract}

\section{Introduction}
\label{sec:introduction}

Interstellar dust grains account for $\simeq$1\% of the mass of the interstellar medium (ISM).  They are involved in different important processes such as photo-electric heating of the neutral interstellar gas, cooling in dense star-forming regions and the formation of molecules, including H$_2$, on their surfaces.  Dust emission can be used to trace the structure of the interstellar medium (ISM) in the Milky Way and in the local Universe (e.g., \cite{foyleetal12,combesetal12,hilletal12}). The thermal dust emission can be modeled using a modified blackbody spectrum in the infrared to submillimeter wavelength range, but physically motivated models remain a subject of debate, since the exact nature of the dust grains is still largely unknown. Understanding dust emission polarization is also important to devise foreground subtraction strategies for CMB experiments. 

ISM dust grains absorb starlight in the visible and ultra-violet, which heats them to temperatures of $\simeq$17\,K in the diffuse ISM in the solar neighborhood in our Galaxy \cite{boulangeretal96}. The polarization of thermal dust emission is believed to arise from the irregular shape of dust grains, and their alignment. The global alignment is believed to be the result of fast grain rotation and relaxation processes slowly bringing the grain minor axis onto the local magnetic field direction (e.g., \cite{Lazarian2003,Lazarian2007}). The grain thermal emission being stronger along the long axis of the grain, the global partial alignment causes a fraction of the thermal emission to be linearly polarized in a direction orthogonal to the magnetic field direction as projected on the sky. For the same reason, non-polarized starlight passing through the ISM with aligned dust grains also becomes polarized, with preferential absorption along the long axis of the grains leading to extinction in the visible and the near-infrared (NIR) being polarized parallel to the magnetic field lines.

%Observations of polarized extinction and emission
First measurements of the polarized extinction in the visible and NIR date from the 1960s (see large catalogs such as in \cite{Mathewson1971}). These studies allowed accurate measurements of the spectral shape of the polarized extinction curve, also known as the Serkowski law (\cite{Serkowski1975}), which is an efficient way of constraining the size distribution of dust grains. Measurements of the thermal dust emission in polarization are more recent. The balloon experiment Archeops (\cite{Benoit2004a}) mapped the polarized dust emission at 353\,GHz with $\sim13^\prime$ resolution over $\sim20$\% of the sky.  These measurements indicated high polarization levels (up to 15\%) in the diffuse ISM. More recently, the Planck satellite mapped the polarized emission over the whole sky in 7 spectral bands in the wavelength range $850\mic$ (353\,GHz) to 1.0\,cm (30\,GHz) \cite{planck2014-xix}. At the highest frequencies, thermal dust  dominates the polarization signal, while low frequencies are typically dominated by polarized synchrotron emission. Analysis of the polarized thermal dust emission at 353\,GHz (\cite{planck2014-xix}) indicated a good correlation with polarized extinction. \cite{planck2014-XXI} showed that the overall thermal dust polarization fraction is only a few percent of the total dust emission over most of the sky, but confirmed the existence of highly polarized regions at high galactic latitudes with polarization fractions reaching up to 22\%. These studies also demonstrated that the thermal dust polarization fraction varies by large factors on small scales. These variations appear linked to the total gas column density, with dense regions exhibiting lower polarization, and to the structure of the magnetic field, with regions showing the most B-field rotation on the plane of the sky also being the least polarized. This latter behavior was shown to be consistent with predictions of MHD models of the ISM (see \cite{planck2014-XX}).
As a consequence of the above studies, the polarized dust thermal emission is now recognized as a dominant foreground contaminant to the observation of the Cosmic Microwave Background (CMB) polarization (see \cite{BICEP2Bmodes}).

%Other Observations in the FIR-submm
Several other facilities allow observations of the thermal dust emission from airborne and ground-based telescopes. The {\HAWCplus} instrument on {\SOFIA} has polarization capabilities in 4 bands from $53\mic$ to $214\mic$ (\cite{Harper2018}). The {\SCUBApol} instrument on {\JCMT} \cite{Holland_etal2013} can also map thermal dust polarization at $850\mic$. The {\NIKAtwo} instrument on the {\IRAM} 30m telescope \cite{Calvo_etal2016} can be used to measure polarization at 260 GHz (1.1 mm) with angular resolution of 10". Finally, the ALMA interferometer allows polarization measurements in band 7 (350 GHz) with very high angular resolution (\cite{Hull2020}). The {\Blastpol} instrument (\cite{Fissel2010}) measures polarized dust emission in 3 bands from $250\mic$ to $500\mic$. In most cases, these facilities are limited in sensitivity to observations of very bright regions and/or suffer from atmospheric absorption or emission fluctuations. Because they can only map fields of view that are limited in size, at much better angular resolution than {\Planck}, a comparison of their results with those of {\Planck} for the same region is at best very difficult, sometimes impossible.

%About the importance of constraining the thermal dust polarization SED
%Correcting for the contribution of the polarized foreground requires knowledge of its spectral and spatial variations.
Measuring the spectral and spatial variations of polarized dust emission provides a potentially powerful constraint on the physics of dust grains (see for instance \cite{guillet2018}), and is crucial to accurately separate the contribution of the polarized Galactic foreground from the CMB signal. To date, spectral variations of dust polarization have been only poorly constrained by observations. \cite{planck2014-XXII} established the first reliable measurement of the spectral variations over the {\Planck} frequency range, using the average dust emission over a carefully selected fraction of the sky. This study concluded that the polarization fraction is roughly constant across 353\,GHz to 100\,GHz, with some indication (at the $3\sigma$ level) that the polarization fraction decreases with decreasing frequency. This measurement of the spectral shape of the dust polarization fraction is extremely challenging due to the decreasing brightness of dust emission at low frequencies and the increasing contribution of polarized synchrotron emission and unpolarized sources such as spinning dust emission and free-free. At frequencies above 353\,GHz, most existing measurements have been obtained by large ground-based or airborne telescopes, which can only map very restricted regions around bright sources. Differences in resolution and the differential scale filtering necessary to subtract atmospheric emission complicate an co-analysis of these measurements and the {\Planck} data.  As a consequence, there is so far very little information available about the polarized SED of thermal dust emission. A key objective of the {\PILOT} mission is to improve our understanding of the thermal dust polarization signal, by measuring it at higher frequencies than {\Planck} in the far-infrared, at an angular resolution and spatial coverage that enables a robust co-analysis with the {\Planck} data.

%About the difficulties associated to systematic effect in polarization
Measurements of astrophysical polarization are difficult because the signal is extremely weak. Most, if not all, of the instruments mentioned above have encountered  difficulties in accurately measuring polarization at low intensities due to systematic instrumental effects. Some of these effects result from well-understood phenomena, such as imperfect inter-calibration of detectors, inaccurate correction for time constants of detectors and for electronic cross-talk, ADC conversion, unmasked glitches, etc. Other systematic effects have been discovered during data processing, such as the spurious contributions from molecular gas spectral lines in the signal \cite{planck2013-p03a} and bandpass mismatch between detectors \cite{Planck2020_III_HFI_DPC_leakage,Banerji2019}, both of which were encountered in the {\Planck} data and required dedicated complex treatment \cite{Lopez_Radcenco2021SRoll3}. Another example is the effect of the Gore-Tex membrane in front of the {\JCMT} which requires special treatment (\cite{Fribergetal2018}). Recently, a leakage from intensity to polarization has been identified by several experiments including {\NIKAtwo} (\cite{Ritacco2017,Ajeddig2020,Ajeddig2022}), and {\HAWCplus} (\cite{LopezRodriguezetal2022}) as a clear limitation to the accuracy of polarization measurement. This effect appears to originate from imperfections of the optical systems that lead to asymmetries in the optical ray propagation through the instrument, producing artificial polarization signal on un-polarized sources. The exact origin is not fully understood and may be instrument dependent.

In this paper, we present the method used to correct for the polarization leakage in the PILOT data and evaluate its performance. We describe two other systematic effects that have an electronic origin, -- detector cross-talk and a readout electronics memory effect -- that affect the PILOT point spread function (PSF) and must be addressed prior to the leakage correction. We give a short description of the instrument and the flights in Sect.\,\ref{sec:pilot_instrument} and Sect.\,\ref{sec:flights} respectively. In Sect.\,\ref{sec:systematic_effects} we present observations of Jupiter obtained during flight 3, which show the effect of {\ghost}, {\crosstalk} and {\leakage}. We use the Jupiter data to characterize and correct for the above systematic effects. We show residual maps to assess the uncertainties associated with residual systematic effects after correction and measure the performance of the leakage correction, in Sect.\,\ref{sec:leakage_performance}. We summarize our conclusions in Sect\,\ref{sec:conclusion}.

\section{The {\Pilot} instrument}
\label{sec:pilot_instrument}

\begin{table}[ht]
\caption{\label{tab:summarize_optics}
Key optical characteristics of the {\PILOT} instrument.}
\begin{center}
\begin{tabular}{|l|c|c|}
\hline\noalign{\smallskip}
\hline
Telescope type & \multicolumn{2}{|c|}{Gregorian} \\
Equivalent focal length [mm] & \multicolumn{2}{|c|}{1790} \\
Numerical aperture & \multicolumn{2}{|c|}{$F/2.5$} \\
FOV [$\degr$] & \multicolumn{2}{|c|}{$1.0 \times 0.8$} \\
Ceiling altitude & \multicolumn{2}{|c|}{$\sim$3 hPa} \\
Pointing reconstruction & \multicolumn{2}{|c|}{translation$=1{\second}$, $1\sigma$}\\
 & \multicolumn{2}{|c|}{rotation$=6{\second}$, $1\sigma$}\\
Gondola mass & \multicolumn{2}{|c|}{$\sim$1100 kg} \\
\hline
\hline
Primary mirror (M1) & \multicolumn{2}{|c|}{Off-axis parabolic}\\
M1 diameter [mm] & \multicolumn{2}{|c|}{$930 \times 830$}\\
M1 used diameter [mm] & \multicolumn{2}{|c|}{730}\\
Focal length [mm] & \multicolumn{2}{|c|}{750} \\
\hline
\hline
Detector type & \multicolumn{2}{|c|}{multiplexed} \\
  & \multicolumn{2}{|c|}{bolometer arrays} \\
Total number of detectors  & \multicolumn{2}{|c|}{2048} \\
Detectors temperature [mK] & \multicolumn{2}{|c|}{300} \\
Sampling rate [Hz] & \multicolumn{2}{|c|}{40}\\
\hline
\hline
Photometric channels &  \multicolumn{2}{|c|}{}  \\
$\lambda_0$ [$\mic$] & \multicolumn{2}{|c|}{$240$} \\
$\nu_0$ [GHz] & \multicolumn{2}{|c|}{$1250$} \\
$\Delta\nu/\nu$ & \multicolumn{2}{|c|}{$0.27$} \\
beam FWHM [{\minute}] & \multicolumn{2}{|c|}{1.9} \\
Minimum Strehl ratio & \multicolumn{2}{|c|}{0.95} \\
\hline
\noalign{\smallskip}\hline
\end{tabular}
\end{center}
\end{table}

A complete description of the {\Pilot} instrument is available in \cite{Bernard_etal2016}.  Here, we only give a brief description for completeness. Table \ref{tab:summarize_optics} summarizes the main characteristics of the instrument.\\

The telescope optics comprises an off-axis paraboloid primary mirror (M1) with diameter of 0.83\,m and an off-axis ellipsoid secondary mirror (M2). The combination respects the Mizuguchi-Dragone condition to minimize depolarization effects (see \cite{Bernard_etal2016,Engel2013}).  All optics following M1, including M2, are cooled to a cryogenic temperature of 2\,K.

Following the Gregorian telescope, the beam is folded using a flat mirror (M3) towards a re-imager and a polarimeter.  Two lenses (L1 and L2) are used to re-image the focus of the telescope on the detectors. A Lyot-stop is placed between the lenses at a pupil plane that is a conjugate of the primary mirror. A rotating Half-Wave Plate ({\HWP}), made of sapphire, is located next to the Lyot-stop.  The bi-refringent material of the {\HWP} introduces a phase shift between the two orthogonal polarization components of the incident light. A polarization analyzer consisting of parallel metallic wires is placed at a $45\degr$ angle in front of the detectors, in order to transmit one polarization to the transmission ({\TRANS}) focal plane and reflect the other polarization to the reflection ({\REFLEX}) focal plane.  Observations at two or greater different {\HWP} angles allow us to reconstruct the Stokes parameters I, Q and U as described in Sect.\,\ref{sec:Polarisation_measurements}.  Each of the {\TRANS} and {\REFLEX} focal planes includes 1024 bolometers (4 arrays of 16 X 16 pixels). They are cooled to 300\,mK  by a closed cycle $^3$He fridge. The detectors were developed by CEA/LETI for the {\Pacs} instrument on board the {\Herschel} satellite.

In order to reconstruct the pointing of the instrument, we use the {\Estadius} stellar sensor developed by CNES for stratospheric applications and described in \cite{Montel+2015}.  This system provides an angular resolution of a few arcseconds, which is required to optimally combine observations of the same part of the sky obtained with various polarization analysis angles. A key feature of {\Estadius} is that it remains accurate even with fast scan speeds (up to  1$\degr$/s).  An internal calibration source ({\ICS}) is used inflight to calibrate time variations of the detector responses. This device is described in \cite{Hargrave2006,Hargrave2003}. The source is located behind mirror M3 and illuminates all detectors simultaneously. It is driven using a square modulated current. The current and voltage of the source are measured continuously during flight, in order to monitor the power dissipated in the source.

\subsection{Polarization measurements}
\label{sec:Polarisation_measurements}

\begin{figure*}[ht]
\begin{center}
\includegraphics[width=0.95\textwidth]{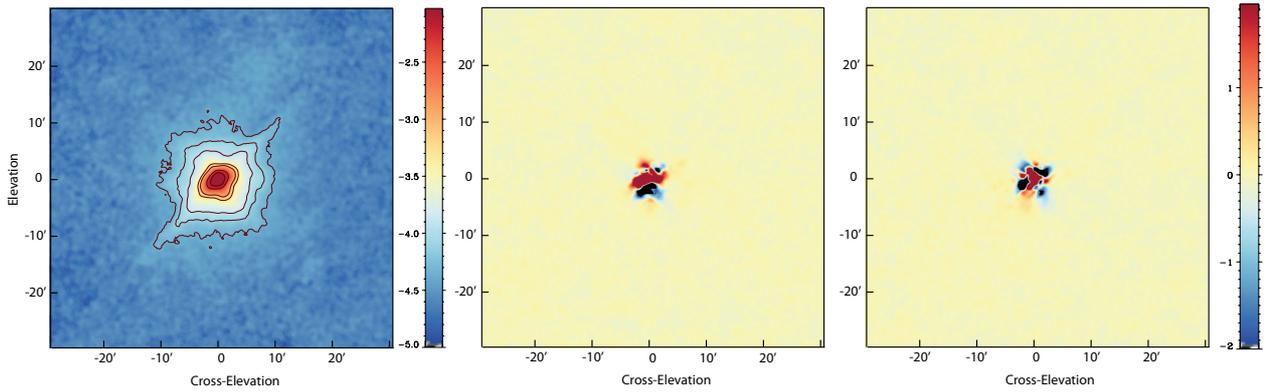}
\caption{\label{fig:leakage_psf} Total intensity $\StokesI$ (left), $\StokesQ$ (center) and $\StokesU$ (right) beam maps obtained on Jupiter during {\flightthree} before correction of systematic effects discussed here. The images are shown in instrument coordinates with elevation increasing upward and cross-elevation (azimuth) increasing to the right. The images are shown in arbitrary units in logarithmic scale for $\StokesI$ and linear scale for  $\StokesQ$ and $\StokesU$. The elongation across the first diagonal in the total intensity image is due to {\crosstalk}. The negative shadow of the {\crosstalk} signal, the PSF distortion visible above the lower left-upper right diagonal and the faint residual source appearing below the planet along the other diagonal are due to {\ghost}. The non-zero $\StokesQ$ and $\StokesU$ originate from intensity to polarization {\leakage}. These effects are described in Sect.\,\ref{sec:systematic_effects}}
\end{center}
\end{figure*}

Assuming a perfect {\HWP}, the {\Pilot} measurements $m$ are related to the input Stokes parameters $\StokesI$, $\StokesQ$, $\StokesU$ of partially linearly polarized light through
\begin{equation}
m = R_{xy} T_{xy} \times [\StokesI \pm \StokesQ_{inst} \cos4\omega \pm \StokesU_{inst} \sin4\omega]+O_{xy},
\label{eq:pol_measure_easy_Stokes}
\end{equation}
where $R_{xy}$ and $T_{xy}$ are the detectors response and optical transmission respectively, and $O_{xy}$ is an arbitrary electronics offset.  For the configuration of the {\HWP} and polarizer in the instrument, $\omega$ is the angle between the {\HWP} fast axis direction and the horizontal direction measured counterclockwise as seen from the instrument. The $\pm$ sign is $+$ and $-$ for the {\REFLEX} and {\TRANS} arrays respectively (see \cite{Bernard_etal2016}). Note that, with the above conventions, $\StokesQ_{inst}$ and $\StokesU_{inst}$ are defined with respect to instrument coordinates in the IAU convention, with $\StokesQ_{inst}$=0 for vertical polarization.
{For {\PILOT}, $\omega$ is related to a mechanical {\HWP} position called $\hwppos$, which can be varied continuously over the range 1 $\le \hwppos \le$ 8 as
\begin{equation}
\omega = 87.25\degr -(\hwppos - 5) \times 11.25\degr,
\label{eq:pol_hwp_angle}
\end{equation}
allowing the {\HWP} fast axis to vary by approx. $\pm 45\degr$ around the vertical direction.}
When referring to the sky polarization $\StokesQ$ and $\StokesU$, Eq.\,\ref{eq:pol_measure_easy_Stokes} becomes
\begin{equation}
m = R_{xy} T_{xy} \times [\StokesI \pm \StokesQ \cos(2\theta) \pm \StokesU \sin(2\theta)]+O_{xy},
\label{eq:pol_measure_easy_Stokes_sky}
\end{equation}
where $\theta=2\times\omega+\phi$ is the analysis angle, $\phi$ is the time varying parallactic angle measured counterclockwise from equatorial north to Zenith for the time and direction of the current observation, and $\StokesQ$ and $\StokesU$ are in the IAU convention with respect to equatorial coordinates. In practice, maps of $\StokesQ$ and $\StokesU$ are derived from observing the same patch of sky with at least two values of the analysis angle taken at different times in general. Inversion to derive sky maps of $\StokesI$, $\StokesQ$ and $\StokesU$ can be done through polarization map-making algorithms (see for instance \cite{deGasperis+2005}). The light polarization fraction $\polfrac$ and polarization direction $\polang$ are then defined as:
\begin{equation}
\label{eq:pem}
\polfrac=\frac{\sqrt{\StokesQ^2+\StokesU^2}}{\StokesI}
\end{equation}
and
\begin{equation}
\label{eq:thetam}
\polang=0.5 \times \arctan(\StokesU/\StokesQ).
\end{equation}

\section{The {\Pilot} flights and observations}
\label{sec:flights}

\begin{table}[ht]
\centering
\caption{\label{tab:flight3-observations} Observations obtained during  {\flightthree}.}
\begin{tabular}{llll}
\hline
\multicolumn{1}{c}{Source} &
\multicolumn{1}{c}{\begin{tabular}[c]{@{}c@{}}Observation Time \\
                     {[}mn{]}\end{tabular}} &
\multicolumn{1}{c}{\begin{tabular}[c]{@{}c@{}}Map size\\ {[}deg x
                     deg{]}\end{tabular}} &
\begin{tabular}[c]{@{}l@{}}Total depth\\ {[}$deg^2/h${]}\end{tabular}  \\
\hline
Aquila Rift                & 128.5      & 7 x 2    & 6.5 \\
Crab nebula                  & 100.      & 1.5 x 1.2    & 1.1 \\
Fan             & 118.5 & 5 x 3.2  & 0.8 \\
Jupiter        & 33. & 2 x 1    & 3.6 \\
M31            & 301.6 & 4 x 1.8    & 1.4 \\
MW L133       & 101.58 & 3 x 2.8 & 5.0 \\
Orion  & 140.1 & 5 x 2.5 & 5.3 \\
Tau B211             & 50.1 & 2 x 1.8 & 4.3 \\
Tau L1506            & 160   & 2 x 1.9    & 1.4 \\
SkydipM31            & 20.   & 32 x 2 & n/a \\
SkydipPol           & 33.1   & 44 x 2   & n/a\\ 
\hline
\end{tabular}
\end{table}

{\Pilot} is carried to the stratosphere by a generic gondola suspended under an open stratospheric balloon through a flight chain, with a helium gas volume of $\sim$ 800\,000~m$^3$ at ceiling altitude. The flights are operated by the French National Space Agency (CNES) with launch campaigns involving several international balloon experiments (up to six per campaign).

At ceiling altitude, the instrument can be pointed towards a given sky direction using the gondola rotation around the flight chain and rotation of the instrument around the elevation axis (see \cite{Bernard_etal2016}). Scientific observations are organized into individual observing tiles (also called observations for short) during which a given rectangular region of the sky is scanned by combining the azimuth and elevation rotations.
The flight plan is built taking into account the various observational constraints such as the visibility of astronomical sources, the minimum allowed angular distance between the instrument optical axis and bright sources such as the Sun or the Moon, elevation limits due to the presence of the Earth at low elevations and the balloon at high elevations. The expected performance of the instrument is taken into account when establishing the flight plan, in order to distribute the observing time according to the science objectives, and to evenly distribute both the polarization analysis directions (angle $\theta$ in Eq.\,\ref{eq:pol_measure_easy_Stokes_sky}) and the scanning directions for any given astronomical target.\\

The first two flights of the {\Pilot} experiment took place from the launch-base facilities at the airports of Timmins (Ontario, Canada) in 2015, and Alice Springs (Australia) in 2017, respectively. A detailed description of the characteristics of these flights and the corresponding observations are presented in \cite{Mangilli2019b}. In this paper, we focus on the instrument performance during the third {\Pilot} flight.

\subsection{Performance during flight\#3}
\label{sec:flightthree}

The third flight of the {\PILOT} instrument took place from Timmins on September 24 2019, as part of a balloon experiment launch campaign led by CNES and the Canadian Space Agency (CSA).

The flight lasted approximately 26~hr, during which 21~hr of scientific observations were obtained. The launch took place at 5:36~AM local time. The experiment reached ceiling altitude about 2.3~hr after launch. The instrument reached an altitude of 39~km, slowly decreasing to 37~km during the first day of the flight. The altitude decreased to 34~km during the night due to the lower buoyancy force of the balloon. During the second day, the altitude rose again, reaching 37.5~km just before the gondola was dropped in Quebec. The temperatures of the focal planes evolved slightly with altitude during the ceiling period and remained in the range $296.5$ to $297$~mK and $300$ to $301$~mK for the TRANS and REFLEX focal planes respectively during the day, and $\simeq 297.5$~mK and $\simeq 301.5$~mK during the night. The higher nocturnal temperatures are due to less efficient pumping on the He bath. Out of the eight bolometer arrays, array~\#1 (TRANS), array~\#5 (TRANS) and array~\#6 (TRANS) were not operational during flight\#3. The footprint of the available arrays on the sky is shown in Fig.\,\ref{fig:leakage_fp_distribution}.

The balloon followed a trajectory towards the north-east during most of the flight. We successfully used the two telemetry antennas located in Timmins and Chibougamau. The gondola was recovered about 900\,km north-east of the launch site, north of Saguenay, Quebec. The gondola was brought back to the Timmins base using a helicopter and a truck. The gondola and the instrument suffered no major damage from landing or recovery, which was later confirmed by a thorough inspection following the return of the instrument to France. The astronomical sources targeted during flight\#3 are listed in Table\,\ref{tab:flight3-observations}. In the following, we concentrate on the analysis of the data obtained on Jupiter, which we use to characterize systematic effects.

\section{Systematic effects}
\label{sec:systematic_effects}

\begin{figure*}[ht]
\begin{center}
\includegraphics[width=0.95\textwidth]{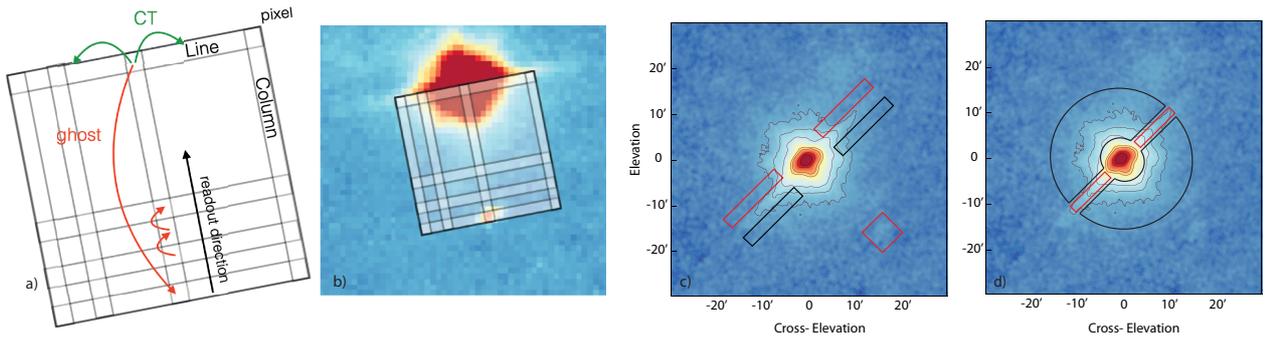}
\caption{
\label{fig:jupiter_sketch}
From left to right: (a):
Schematic view of the array showing the readout direction and the effect of {\crosstalk} along lines and {\ghost} effects.
(b): Image of Jupiter obtained by leaving in glitch detected samples, overlaid with the footprint of the detectors, showing the effect of the {\crosstalk} along the lines and the presence of a ghost opposite to the planet along readout columns due to {\ghost}.
(c): The regions overlaid on the image of Jupiter delineate
the zones used to measure the {\ghost} parameters (red) and the corresponding reference regions (black).
(d): The regions overlaid on the image of Jupiter delineate
the zones used to measure the {\crosstalk} effect parameters (red) and the corresponding reference regions (black).
}
\end{center}
\end{figure*}

In this section we describe three instrumental systematic effects of the {\PILOT} data, not addressed in \cite{Mangilli2019a}. Two of these effects are related to the readout electronics of the {\PILOT} detectors, which we refer to as {\crosstalk} and {\ghost}. The third effect is produced by the optics of the instrument, which we refer to as {\leakage} (see Sect.\,\ref{sec:leakage_characterisation}). Here, we describe the manifestation of each of these effects on the instrument Point Spread Function (PSF) as observed on Jupiter during flight\#3, how we measured the parameters used in the correction methods, how the corrections were performed and the overall performance of the corrections, as measured on the Jupiter observations.

During each flight of the instrument, we observed planets, which can be considered point sources at the resolution of {\Pilot}. These observations can be used to assess the optical quality through a measurement of the PSF. During Flight\#3, we observed Jupiter at its maximum elevation of $\simeq 17\degree$ during about 30\,min at the start of the flight. We obtained eight individual maps using eight positions of the {\HWP}, sampling the available analysis range uniformly. The maps were obtained at two different scanning angles to enable low frequency noise removal. The data were corrected for the responses calculated on the residual atmospheric signal and the {\ICS} calibration signals, and corrected for the effect of the detector time constants through deconvolution, as described in \cite{Mangilli2019a}. The signal was then processed using the {\scanamorphos} map-making software described in \cite{Roussel2013} and as used in its polarization version in \cite{Mangilli2019b} to produce maps of the Stokes parameters {\StokesI}, {\StokesQ} and {\StokesU} and the corresponding variances and co-variances. Note that these maps can also be obtained in instrument coordinates, also referred to as cross-elevation and elevation, by setting the parallactic angle to zero in Eq.\,\ref{eq:pol_measure_easy_Stokes_sky}. This representation is optimal to reveal and characterize effects that project in the focal plane of the instrument, since it avoids blurring through sky rotation.

In order to produce PSF maps that are sufficiently accurate to be used for {\leakage} correction, we constructed Jupiter maps with a pixel size of $6\second$. We generated individual maps for each detector array and for each individual observation of the planet. We use these maps for assessing potential temporal or focal plane variations of the systematic effects.

Figure\,\ref{fig:leakage_psf} shows the maps of Jupiter in instrumental coordinates where all the data from flight\#3 have been used. Below, we use these maps to investigate systematic effects affecting the PSF in polarization, and to measure the parameters used in the correction method. 

\subsection{{\Crosstalk}}
\label{sec:crosstalk}

Figure\,\ref{fig:leakage_psf} shows the total intensity map of Jupiter obtained during flight\#3 in instrument coordinates. A blurred linear extension is clearly visible across the PSF from the lower-left to upper-right of the image. As illustrated in Fig.\,\ref{fig:jupiter_sketch}, this direction corresponds to the orientation of the individual pixel lines on the arrays, which are rotated $45\degree$ with respect to instrument coordinates. The readout electronics of the {\PILOT} detectors is such that the signals from bolometers along each line are transferred simultaneously to a buffer unit (BU) with 16 registers for amplification. We interpret the observed effect as cross-talk between pixels along a given detector line, which could happen within the BU. As the strong signal from the peak of the optical PSF falls on a given pixel of the array, part of its intensity is transferred through {\crosstalk} to other pixels along the corresponding detector line, producing the observed pattern. This is illustrated in Fig.\,\ref{fig:jupiter_sketch} which shows the projection of the array footprint on the Jupiter map.

We model the {\crosstalk} as the transfer of a fraction $\fct(i,\iprime)$ of the signal from pixel $i$ to pixel $\iprime$ along line $j$. As the transfer is likely to occur in the buffer unit which is common to all lines, we further assume that $\fct(i,\iprime)$ is independent of line number $j$. We therefore subtract {\crosstalk} following
\begin{equation}
\label{eq:ct_correction}
d^\prime_{ij}=d_{ij}
- \sum_{\iprime \neq i}^{}{\fct(i,\iprime) \times d_{\iprime j}}
+ \sum_{\iprime \neq i}^{}{\fct(i,\iprime) \times d_{ij}} ,
\end{equation}
where $d_{ij}$ and $d^\prime_{ij}$ are the signal before and after {\crosstalk} subtraction respectively, and the summations are carried out over all pixels along the line under consideration. The two terms correspond to the signal received and given by the considered pixel. We also assume that, {\crosstalk} being an induction effect, it is symmetric with $\fct(i,\iprime)$ = $\fct(\iprime,i)$.

%glitch analysis to see if there are variations of ct factor
We searched for possible variations of the {\crosstalk} coefficient along detector lines. For this, we analyzed jointly the recordings of pixels receiving a strong glitch (normally masked out during the processing) and the stacked signal of simultaneous recordings of other pixels of the same line.
We correlated the stacked signals from the main glitched pixel and the cross-talk pixel. This analysis did not produce convincing evidence for variations of $\fct(i,\iprime)$ along the pixel lines. We therefore assume that $\fct(i,\iprime)$ does not vary across a given array and we search for a single value of the pixel-to-pixel cross-talk coefficient $\fct$ for each array.

%How we measure cross-talk
In order to determine $\fct$ for each array, we defined a cross-talk region and a reference region in each image of Jupiter (see Fig.\,\ref{fig:jupiter_sketch}). Both regions share the same average distance from the planet so that they would have the same brightness in the absence of {\crosstalk} but are centered on a regions of high and low cross-talk signals respectively. The value of $\fct$ for each array was found through a $\chi^2$ minimization of the signal difference between the cross-talk and the reference regions in images of Jupiter obtained with each array separately. At each iteration of the minimization, the cross-talk signal was subtracted from the timeline using Eq.\,\ref{eq:ct_correction} and a new image was produced. The resulting values of the {\crosstalk} parameters $\fct$ are given in Tab.\,\ref{tab:crosstalk_params}. The derived values appear consistent between arrays and at a level just below $10^{-3}$. The values are comparable between arrays and there appear to be no particular similarities between parameter values for arrays sharing the same BU.

The Jupiter map after correction of the {\crosstalk} using the parameters given in Tab.\,\ref{tab:crosstalk_params} and Eq.\,\ref{eq:ct_correction} is shown in Fig.\,\ref{fig:jupiter_corrections}.

\begin{table}[ht]
\centering
\caption{\label{tab:crosstalk_params} Cross-talk parameters derived for each array. Column 1 gives the name of the corresponding focal plane, column 2 gives the array number, column 3 gives the Buffer Unit (BU) number associated to reading the array, column 4 gives the value for the {\crosstalk} coefficient derived.}
\begin{tabular}{llll}
\hline
Focal Plane & Array & BU & $\fct$  \\
\hline
TRANS & 2 & 2 & $9.08 10^{-4}$  \\
\hline
REFLEX & 3 & 3 & $8.31 10^{-4}$  \\
REFLEX & 4 & 3 & $9.08 10^{-4}$ \\
\hline
REFLEX & 7 & 4 & $9.69 10^{-4}$ \\
REFLEX & 8 & 4 & $9.33 10^{-4}$ \\
\hline
\end{tabular}
\end{table}

Following the correction for {\crosstalk}, the amplitude of the effect as measured in the regions defined above is about $0.9\%$ of the PSF peak value, significantly smaller than the initial value of $4.8\%$.

\subsection{{\Ghost}}
\label{sec:ghost_characterisation}

\begin{figure*}[ht]
\begin{center}
\includegraphics[width=0.9\textwidth]{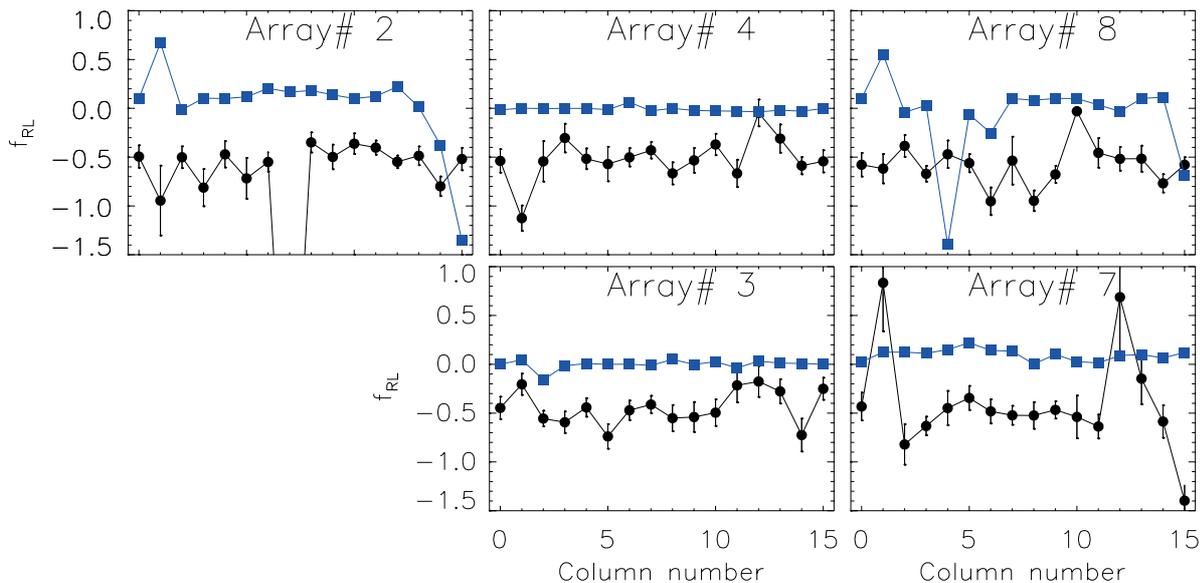}
\caption{\label{fig:ghost_params}
Parameters for the {\ghost} derived for each array, plotted as a function of the readout column. The black curve shows the parameter values between adjacent pixels along the column. The blue curve shows the parameter values when readout returns from line 16 to line 1.
}
\end{center}
\end{figure*}

Figure\,\ref{fig:jupiter_sketch} shows a map of Jupiter obtained during {\flightthree} where the {\scanamorphos} glitch detection was inhibited during processing. The map is overlaid with the footprint of one of the {\PILOT} bolometer arrays. The image clearly shows some positive signal located precisely one array away from Jupiter along the column direction of the array. This fake source appears in the data stream of bolometers located on line 1 of each array, only when a bright source is present on the same column on line 16. This effect had already been seen clearly in calibration data, when a bright source was moved over all pixels of all arrays (\cite{Misawa_etal2017}). It is attributed to latency in the time-multiplexed readout electronics, which we refer to as {\ghost}. This effect transfers some of the signal from one readout to the next along the readout direction (see Fig.\,\ref{fig:jupiter_sketch}), including when the readout goes from line 16 back to line 1, which creates the fake positive source in the map. The effect is also seen as a negative shadow of the {\crosstalk} signal described in Sect.\,\ref{sec:crosstalk}, which indicates that the {\ghost} effect is mostly negative during transfer across the array and positive when readout is reset to line 1. The fact that we see the effect of the {\ghost} on the {\crosstalk} signal also shows that the {\ghost} happens after the {\crosstalk} in the detection chain, and as a consequence it needs to be corrected before {\crosstalk} in the data processing.

We model the {\ghost} effect as the transfer of a fraction $\frl(i,j)$ of the signal from readout $j$ to readout $j+1$ along a given column $i$. We correct for the effect in the timelines using
\begin{equation}
\label{eq:rf_correction}
d^\prime_{ij}=d_{ij}
- \frl(i,j-1) \times d_{ij-1}
+ \frl(i,j+1) \times d_{ij+1},
\end{equation}
where all readouts have been ordered with time of acquisition. The first term corresponds to the signal received by the considered pixel from the previous readout and the second term corresponds to the signal given to the next readout. 
We further assume the same value for $\frl(i,j)$ between all consecutive readouts, except for multiples of 16 ($j=n\times 16$) with value $\frl(i,16)$ in Eq.\,\ref{eq:rf_correction}.
In order to measure the parameter $\frl(i,j)$ for each readout column $i$ and array for $j \ne n \times 16$ , we constructed images of Jupiter in instrument coordinates using only the signal from that individual column of that individual array, using a timeline corrected according to Eq.\,\ref{eq:rf_correction}. We optimized the value of $\frl(i,j)$ in order to minimize the difference between the average signal in two rectangular boxes located on both sides of the {\crosstalk} extension around Jupiter, as shown in Fig.\,\ref{fig:jupiter_sketch}.
In order to measure $\frl(i,16)$, we performed a similar minimization but minimizing the signal in a square box centered on the fake source in the images as shown in Fig.\,\ref{fig:jupiter_sketch}.
In both cases, the minimization was performed using the $\chi^2$ minimization algorithm implemented in the $IDL$ routine {\it mpfitfun}.

The values derived for $\frl(i,j)$ are shown in Fig.\,\ref{fig:ghost_params} for each column of each array. The values are generally negative, while 
$\frl(i,16)$ is generally much smaller in absolute value but mostly positive.

\begin{figure*}[ht]
\begin{center}
\includegraphics[width=0.95\textwidth]{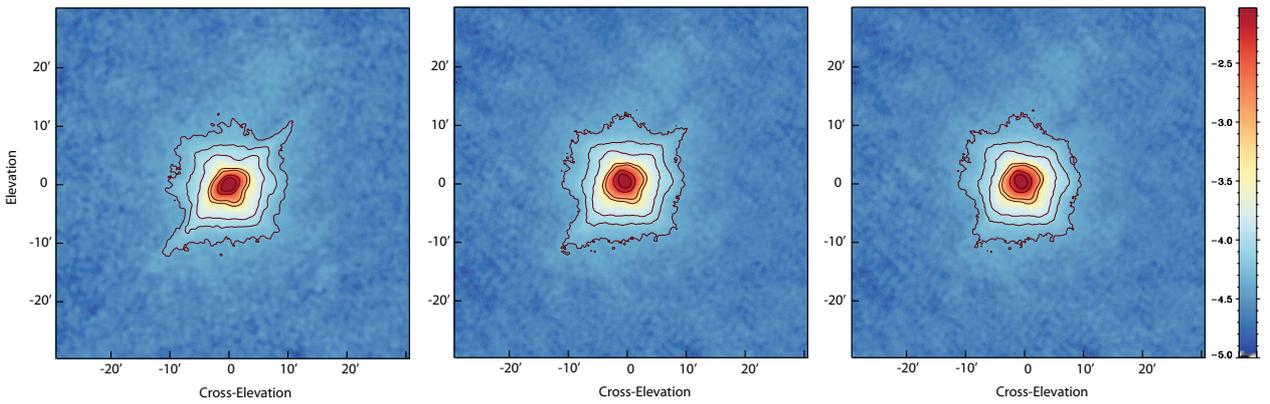}
\caption{\label{fig:jupiter_corrections} Images of Jupiter constructed from uncorrected data (left), corrected for {\ghost} (middle), corrected for {\ghost} and cross-talk (right). All images are shown in the same arbitrary units in logarithmic scale.}
\end{center}
\end{figure*}

The Jupiter map after correction of the {\ghost} using the parameters shown in Fig.\,\ref{fig:ghost_params} and Eq.\,\ref{eq:rf_correction} is shown on Fig.\,\ref{fig:jupiter_corrections}. Note that this correction produces a significant shift of the planet peak position. Since we compute sky coordinates of each data sample using the data from the {\Estadius} stellar sensor and correcting for the offset between the sensor and the {\PILOT} instrument optical axis using the position of observed bright sources, we recompute coordinates following that correction, which we use for the rest of the data processing and analysis. 

\subsection{{\Leakage}}
\label{sec:leakage_characterisation}

\begin{figure}[ht]
\begin{center}
\includegraphics[width=0.45\textwidth]{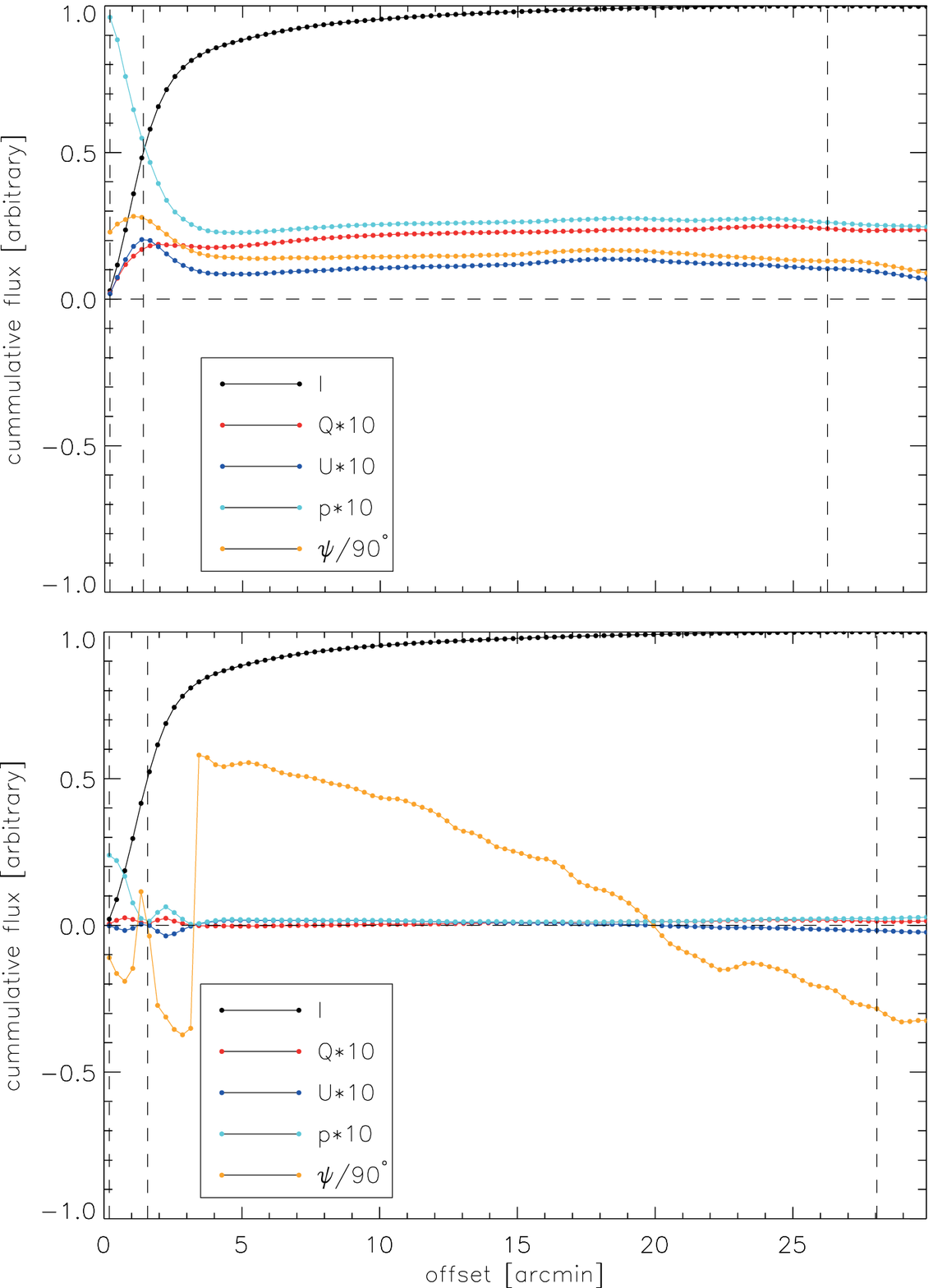}
\caption{\label{fig:leakage_psf_profile_corrected}
Cumulative intensity profiles as a function of integration radius from the peak of the total intensity in Jupiter maps. The black curve shows the cumulative profile of the total intensity. The red and dark-blue curves show the {\StokesQ} and {\StokesU} profiles multiplied by 10. The light blue and yellow curves show the corresponding average polarization leakage fraction in units of 10\%, and the corresponding polarization angle divided by $90\degr$. The top panel shows the initial leakage characteristics before correction. The bottom panel shows the leakage characteristics after correction for {\ghost}, {\crosstalk} and leakage. The erratic behavior of the polarization angle on this panel is due to the very low magnitude of $\StokesQ$ and $\StokesU$ after correction.
}
\end{center}
\end{figure}

\begin{figure}[ht]
\begin{center}
\includegraphics[width=0.5\textwidth]{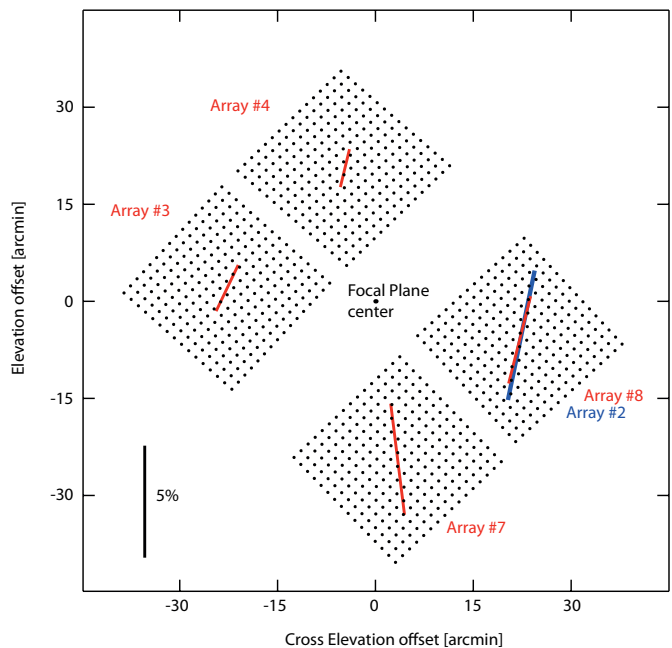}
\caption{\label{fig:leakage_fp_distribution} Distribution of the leakage polarization direction and intensity across the {\PILOT} focal plane.
The focal plane is shown in instrument coordinates, with elevation corresponding to the vertical direction, in offsets from the focal plane center. The circles show the positions of individual pixels of the TRANS arrays. The labels give the name of each operational array in red and blue for the REFLEX and TRANS focal plane respectively. The lines show the direction and amplitude of the intensity to the polarization leakage as measured on Jupiter for each array independently, in red and blue for the REFLEX and TRANS arrays  respectively. The scale on the bottom left shows 5\%.}
\end{center}
\end{figure}

\begin{figure*}[ht]
\begin{center}
\includegraphics[width=0.95\textwidth]{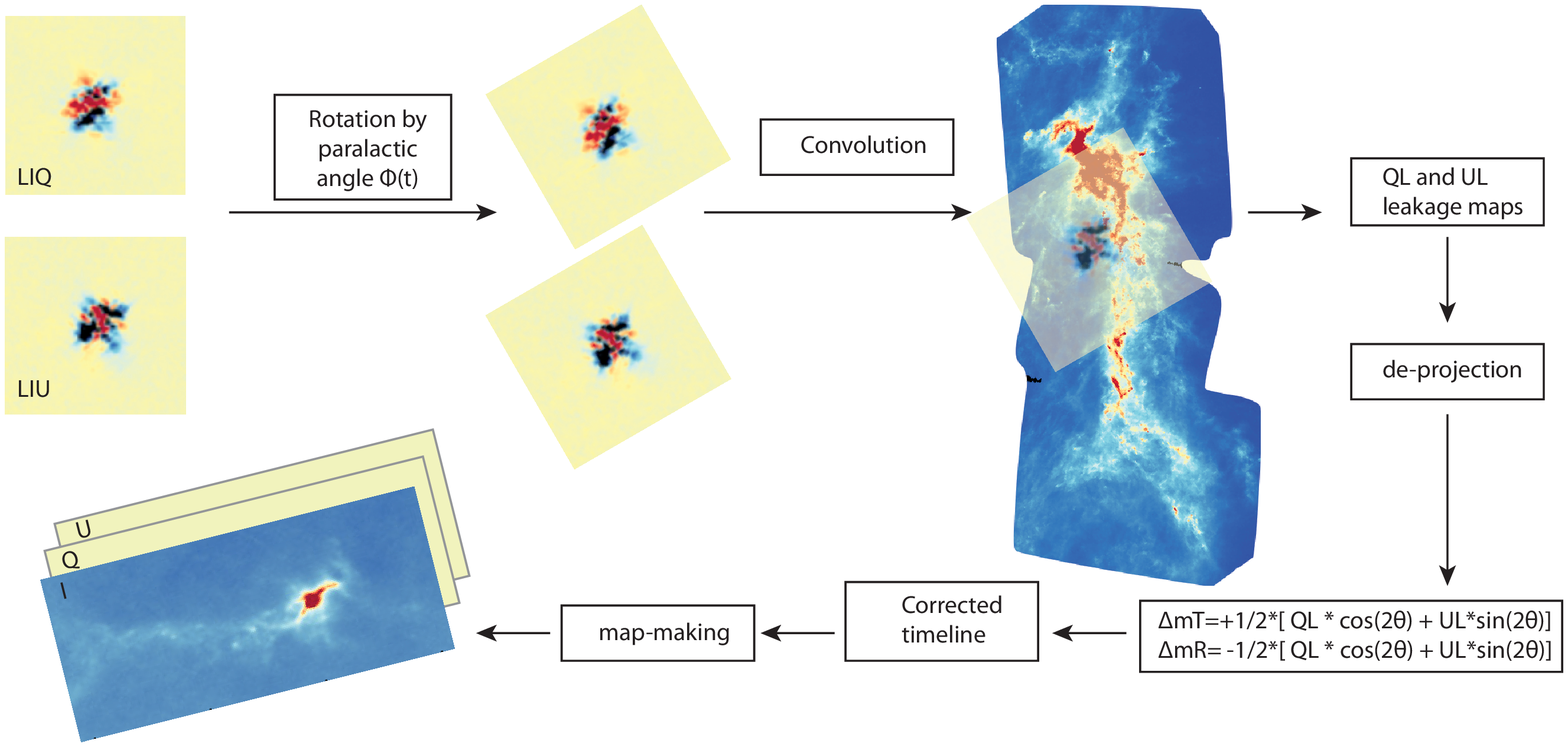}
\caption{\label{fig:leakage_correction} Schematic representation of the method used to correct for the intensity to polarization leakage in the {\PILOT} data. The data processing steps proceed clockwise from top-left. The {\StokesI}, {\StokesQ} and {\StokesU} PSF as measured on Jupiter are rotated to follow the time-dependent sky rotation. They are then convolved to the high resolution {\Herschel} satellite sky map of the considered object (here the Orion~A region) at $250\mic$ sky to produce instantaneous maps of the {\StokesQ} and {\StokesU} leakage. The instantaneous maps are re-observed to produce a correction time-line which is subtracted from the original data. Our final leakage-corrected, science-ready maps are constructed by applying our map-making algorithm to this corrected timeline.
}
\end{center}
\end{figure*}

Figure\,\ref{fig:leakage_psf} shows the $\StokesI$, $\StokesQ$ and $\StokesU$ maps obtained on Jupiter during {\flightthree} projected in instrument coordinates (elevation and cross-elevation). While we expect Jupiter to show no polarization, the maps clearly show some residual $\StokesQ$ and $\StokesU$ structures at the location of the planet.

Figure\,\ref{fig:leakage_psf_profile_corrected} shows the cumulative profiles of the $\StokesI$, $\StokesQ$ and $\StokesU$ as a function of the integration radius in the maps of Fig.\,\ref{fig:leakage_psf}, as well as the corresponding profiles for polarization fraction {\polfrac} and polarization angle $\polang$.
It is clear that even when averaged over a large area, the $\StokesQ$ and $\StokesU$ cumulative profiles do not converge to zero.
The corresponding values of the leakage are summarized in Table\,\ref{tab:leakage_perfo}. Before leakage correction, the leakage polarization is of the order of {\polfrac}=2.4\% when integrated over the whole PSF and {\polfrac}=4.7\% when integrated over the PSF down to its FWHM. This is non-negligible compared to the typical polarization of thermal dust in the ISM, with a most likely value of around $3\%$, as measured by {\Planck} at 353\,GHz (\cite{planck2014-xix}). Given the amplitude of this leakage effect compared to the expected astrophysical signal, it clearly needs to be subtracted accurately from the data.

This effect is known as instrumental polarization and is also often referred to as  leakage from total intensity to polarization. It has been observed in the polarization data of many, if not all, instruments measuring polarization. Some procedures have been proposed to subtract the effect from the data, for instruments such as  {\NIKAtwo} (\cite{Ritacco2017}), {\ACTPol} (\cite{Guan2021}) or {\HAWCplus} (\cite{LopezRodriguezetal2022}). The origin of the effect is unclear, but observations suggest that it is due to the propagation of light in the instrument.

\begin{table}[ht]
\centering
\caption{\label{tab:leakage_perfo} Leakage correction performance. Columns 1 to 3 indicate which corrections are applied. Column 4 gives the polarization fraction of the leakage as integrated over the whole PSF. Column 5 gives polarization fraction of the leakage as integrated up to the FWHM of the PSF.}
\begin{tabular}{lllll}
\hline
RL & CT & leakage & $p_{fwhm} [\%]$ & $p_{tot} [\%]$  \\
\hline
- & - & - & 5.36 & 2.19 \\    %SC110013_JUPITER_V2p1_PRO_5000000010
x & - & - & 5.38 & 2.62  \\   %SC110011_JUPITER_V2p1_PRO_5000001010
x & x & - & 4.74 & 2.41 \\    %SC110013_JUPITER_V2p2_PRO_5000001210
x & x & x & 0.17 & 0.28   \\  %SC110013_JUPITER_V2p2_PRO_5000001211
\hline
\end{tabular}
\end{table}

As seen in Fig.\,\ref{fig:leakage_psf}, the leakage pattern of the {\PILOT} instrument does not show any distinctive structure. This is unlike the pattern observed for the {\NIKAtwo} instrument for instance. This difference may be due to the fact that we use an off-axis telescope that has no occultation by the support for the secondary mirror. The leakage is also seen to produce mostly $\StokesU$ in instrument coordinates, which indicates that the instrumental polarization is mostly horizontal in those coordinates, corresponding to a polarization vector roughly vertical. This is compatible with the leakage being due to asymmetries in the optical system, which is mostly symmetrical with respect to the vertical direction for the {\PILOT} instrument (see for instance \cite{Bernard_etal2016}).

%spatial variations in the focal plane
In order to investigate possible variations of the leakage with the position in the focal plane, we constructed separate maps of Jupiter with the 5 operational arrays available during {\flightthree}. For each image, we computed the integrated leakage polarization fraction and angle integrated over the FWHM of the total intensity beam. Figure\,\ref{fig:leakage_fp_distribution} shows the distribution of the polarization fraction and orientations in the {\PILOT} focal plane. The direction is roughly vertical across the focal plane and the fraction also varies slightly. Note that the directions are consistent between arrays 2 and 8 which are optical conjugates on the sky.

\subsection{{\Leakage} correction performances}
\label{sec:leakage_performance}

%method for Correcting the leakage
The scheme we use to subtract the leakage is illustrated in Fig.\,\ref{fig:leakage_correction}. We adopt the description proposed by \cite{Ritacco2017} for the {\NIKAtwo} data, in which the leakage can be computed as the convolution of the total intensity map of the sky with the leakage PSF measured on a planet. The method implemented in the {\PILOT} pipeline involves rotating the {\StokesQ} and {\StokesU} leakage PSFs by the parallactic angle to bring them to the correct sky orientation. The rotated leakage PSF maps are then convolved with the total intensity map of the sky to produce some leakage {\StokesQ} and {\StokesU} maps. Note that this has to be done at each time sample of the time-line to account for the continuous sky rotation. The leakage {\StokesQ} and {\StokesU} maps are then interpolated at the sky coordinates corresponding to each data sample, in order to predict a leakage signal through Eq.\,\ref{eq:pol_measure_easy_Stokes}. That signal is then subtracted from the original timeline and the corrected timeline is used to produce a map corrected for the leakage, using the {\scanamorphos} map-making algorithm. Note that, unlike in \cite{Ritacco2017}, we do not subtract any fixed polarization contribution other than the above leakage. The performances quoted therefore reflect the correction of the intensity to polarization leakage through the algorithm described here.

%This is how we implement it in details
%==== what we use for the PSF
For the leakage PSF maps, we use the Jupiter maps shown in Fig.\,\ref{fig:leakage_psf} computed in instrument coordinates and with {\StokesQ} and {\StokesU} also in reference to instrumental coordinates. As a consequence, we set the parallactic angle $\phi$ to zero in Eq.\,\ref{eq:pol_measure_easy_Stokes} when computing the leakage contribution to be subtracted from the original timeline.
%==== what we use for the total intensity
For the total intensity map, we use the {\Herschel} map of the astronomical object extracted from the ESA {\Herschel} Science Archive \footnote{http://archives.esac.esa.int/hsa/whsa/}, which we reproject into the appropriate equatorial  coordinates. This approach is preferred to deconvolving our own intensity map from our total intensity PSF, given the accuracy required for a proper subtraction of the leakage signal.
In the case of Jupiter, we use a fake source map where the total intensity map of the planet is computed as a Gaussian with FWHM of $18.1 \second$, i.e. the FWHM of the $250 \mic$ {\Herschel} band. In both cases, we correlate the observed total intensity {\PILOT} map of the object with the {\Herschel} map and use the linear scale factor between the two images to rescale the {\Herschel} map prior to using it to predict the leakage.
%=== how we discretize calculation of leakage maps
In practice, the calculation of the leakage maps is performed at discrete parallactic angle values covering the range for each {\Pilot} observation tile, with a discretization step of $\simeq1\degree$ and the de-projected time-lines are interpolated at the actual parallactic angle value of each data sample from those maps, using linear interpolation.
%=== how we take into account spatial variations of the leakage
The above processing is computed independently for each detector array and we use the PSF of each array as measured in maps of Jupiter computed for that array only. To first order, the processing therefore takes into account the focal-plane variations of the leakage shown in Fig.\,\ref{fig:leakage_fp_distribution}. 

%performance of the Correction (including sensitivity to the details of implementation).
The residual polarization after leakage correction on the Jupiter data is shown in the bottom panel of Fig.\,\ref{fig:leakage_psf_profile_corrected}. It can be seen that the polarization leakage is strongly reduced compared to the profiles prior to correction shown in the upper panel. The residual polarization leakage as measured at the FWHM radius of the PSF and integrated over the whole PSF are given in tab.\,\ref{tab:leakage_perfo} and are respectively $0.17\%$ and $0.38\%$. The spatial distribution of those residuals in the image shown on the radial profiles is likely due to uncertainties in the pointing reconstruction or in the PSF shape, discretization steps used for map rotations, etc\ldots	 We also stress that a high level of accuracy must be preserved at each step of the leakage correction process, e.g. during the map rotations, convolution and timeline de-projection. Most common reprojection routines do not guarantee sufficient accuracy, and we were obliged to use drizzling methods \cite{Paradis2012} at each step. It also required working with map pixels of $6\second$ consistent with the {\Herschel} resolution, which are $\simeq 22$ times smaller than the {\PILOT} beam, to produce acceptable accuracy of the correction. We note that, due to filtering of large scale emission inherent to measurements with bolometers, the leakage on extended sources should be lower than measured here on a point-like source. We consider that the range of performances given in Tab.\,\ref{tab:leakage_perfo} reflect those attainable on astrophysical sources with the current leakage subtraction method.

\section{Conclusions}
\label{sec:conclusion}

In this paper, we have presented the methods used to correct for residual systematic effects in the {\PILOT} data, in addition to those already described in \cite{Mangilli2019a}. In particular, we describe the {\crosstalk} and {\ghost} systematic effects that affect the shape of the instrument PSF in total intensity. We also described how we measure and correct for the intensity to polarization leakage and the method we use to subtract it from the data. The {\crosstalk} and {\ghost} effects are observed in the total intensity maps of Jupiter obtained during the third flight of the {\PILOT} instrument as distortions of the instrument PSF. We measured the parameters characterizing those effects by using a simple pixel-to-pixel transfer model and derived the transfer coefficients by minimizing the PSF defects in the images of Jupiter. Our analysis showed that the {\ghost} effect is observed on the {\crosstalk} signal, indicating that it arises after {\crosstalk} in the detection chain and therefore needs to be corrected first.\\

Following the above correction, images of Jupiter in polarization show polarized signal at the level of $\simeq 3\%$, which we interpret as leakage from total intensity to polarization, also known as instrumental polarization. Polarization leakage likely affects most instruments measuring polarization in the FIR/submm at a similar level. Using images of Jupiter obtained separately for the five bolometer arrays that were operational during flight\#3, we showed that the leakage is mostly oriented parallel to the symmetry axis of the instrument, which points towards residual asymmetries of the optics as the origin of the leakage. We also showed that the polarization direction and fraction vary across the focal plane, an effect that we also take into account in the correction.
We correct for the leakage in the {\PILOT} pipeline following the method initially proposed by \cite{Ritacco2017} to correct for the leakage in the {\NIKAtwo} instrument. We use the {\StokesI}, {\StokesQ} and {\StokesU} PSFs measured on Jupiter, rotated to follow sky rotation and convolved with a scaled intensity sky map obtained by the {\Herschel} satellite at $250\mic$ to predict maps of the leakage in sky coordinates. Those convolved leakage maps are de-projected onto the timeline of each detector to infer the correction for each detector, taking into account at first order the observed spatial variations of the leakage. We emphasize that accuracy must be preserved at each step of the process in the map rotations, convolution and timeline de-projection. This is not guaranteed by most reprojection routines. Applying our correction strategy to the Jupiter data and using a simple synthetic PSF model for the planet yields a residual polarization fraction lower than $\simeq 0.4\%$, which we regard as the accuracy of our leakage correction method.

\begin{acknowledgements}
{\PILOT} is an international project that involves several European institutes.  The institutes that have contributed to hardware developments for {\PILOT} are IRAP and CNES in Toulouse (France), IAS in Orsay (France), CEA in Saclay (France), Sapienza University in Rome (Italy), Cardiff University (UK) and the Scientific Support Office at ESTEC (NL).  This work was supported by the CNES. It is based on the {\Pilot} data obtained during three flight campaigns operated by CNES, under the agreement between CNES and CNRS / INSU. It benefited from the active participation of CNES, IRAP, IAS, CEA and ESTEC to the flight campaigns. This work was supported by the Programme National ``Physique et Chimie du Milieu Interstellaire'' (PCMI) of CNRS / INSU with INC/INP co-funded by CEA and CNES. DA acknowledges SC MES RK grant No. AP08855858 and Nazarbayev University Grant Programme  No. 110119FD4503. This work was supported by World Premier International Research Center Initiative (WPI), MEXT, Japan.

\end{acknowledgements}

\bibliographystyle{spphys}       
%\bibliography{perfo}   
\bibliography{PILOT_leakage}   

\end{document}